# Crystal Oscillators in OSNMA-Enabled Receivers: An Implementation View for Automotive Applications

Francesco Ardizzon[1], Nicola Laurenti
*Department of Information Engineering*
University of Padova
Padova, Italy,
Email: {ardizzonfr, nil}@dei.unipd

Carlo Sarto, Giovanni Gamba
*Qascom S.r.l*
Bassano del Grappa, Vicenza Italy
Email: {carlo.sarto, giovanni.gamba}@qascom.it

Cillian O'Driscoll
*Cillian O'Driscoll Consulting*
Cork, Ireland
Email: cillian@codc.ie

Ignacio Fernandez-Hernandez
*European Commission*
Brussels, Belgium
email:ignacio.fernandez-hernandez@ec.europa.eu

**Abstract**

To ensure the authenticity of navigation data, Galileo Open Service navigation message authentication (OSNMA) requires loose synchronization between the receiver clock and the system time. This means that during the period between clock calibrations, the receiver clock error needs to be smaller than a pre-defined threshold, currently up to 165s for OSNMA. On the other hand relying on the PVT solution to steer the receiver clock or correct its bias may not be possible since this would depend on the very same signals we intend to authenticate. The aim of this work is to investigate the causes of the frequency accuracy loss leading to clock errors and to build a model that, from the datasheet of a real-time clock (RTC) device, allows to bound the error clock during a certain period. The model's main contributors are temperature changes, long-term aging, and offset at calibration, but it includes other factors. We then apply the model to several RTCs from different manufacturers and bound the maximum error for certain periods, with focus on the two-year between-calibration period expected for the smart tachograph, an automotive application which will integrate OSNMA.

## I. INTRODUCTION

In order to ensure the authenticity and the integrity of the transmitted messages, the Timed Efficient Stream Loss-tolerant Authentication (TESLA) (Perrig, et al., 2002) protocol for broadcast authentication requires a loose time synchronization between transmitter and receiver, that is an upper-bound to the time offset between their clocks (Fernandez-Hernandez, et al., 2020a, Fernandez-Hernandez, et al. 2020b, Cucchi, et al., 2021). TESLA is at the core of the Open Service navigation message authentication (OSNMA) protocol, a global navigation satellite system (GNSS) authentication protocol used for the Galileo Open Service (Fernández-Hernández, et al., 2016).

In the context of the OSNMA protocol, it is customary to assume that on the system side the transmission is synchronous since the satellites are equipped with high precision atomic clocks, the drift of which is assumed negligible with respect to that at the receiver. However, on the receiver side, commercial clocks are much less accurate and stable, which lead to a substantial time mismatch between the system and the receiver clocks that accumulates over time, if not corrected.

Typical GNSS receivers correct this error when calculating the PVT (position, velocity and timing) solution. However, for OSNMA, the estimated clock bias from the GNSS-based PVT cannot be trusted since this would mean to rely on the very same signal that we want to authenticate, potentially introducing security vulnerabilities.

In order to limit the impact of the accumulated clock mismatch on OSNMA operation, receiver clocks may re-calibrate from an external, non-GNSS source. For example, the EU smart tachograph (Baldini, et al., 2018) shall incorporate OSNMA on board its vehicles, and clocks can be reset in periodic workshop visits, as shown in the PATROL (Position Authenticated TachogRaph for OSNMA Launch) project (EUSPA, 2024). However, it is necessary to guarantee that the clock mismatch satisfies the OSNMA constraint at all times between successive workshop resets, the so-called holdover period, and through all possible operating conditions in order to ensure constant authenticity of the navigation message.

Aside from OSNMA, loss receiver synchronization is common for other GNSS delayed-disclosure authentication protocols. For instance, the assisted commercial authentication service (ACAS) (Fernandez-Hernandez, et al., 2023), recently proposed for

---

[1] Corresponding Author: Ardizzon Francesco, ardizzonfr@dei.unipd.it

Galileo, or GPS's CHIMERA (Anderson, et al., 2017). Thus, while in this paper we will mainly target OSNMA, the same analysis can be extended to different authentication services.

In the literature many works tackle the problem of device synchronization, for example in the wireless sensor networks (WSN) (Prakash & Kendall, 2010, Djenouri & Bagaa, 2016, Tirado-Andrés & Araujo, 2019). However, most of these works discuss the protocols that should be employed to synchronize the devices and not the actual clock sources. On the other hand, when developing a model for the clock performance losses, most literature focuses on atomic clocks and is often based on the well-known Allan variance (Allan, 1966), although other approaches are also studied, as e.g., (Galleani, et al., 2003). Only few works focus on clock error estimation and prediction in mass market oscillators, e.g., (Giorgi, 2015). A frequent limitation is that, considering the sophisticated models for the mass market clock synchronization, it is difficult to retrieve accurate estimation of the parameters needed to initialize the models. In WSN, IoT devices typically need only millisecond-level precision, thus they usually rely on the network time protocol (NTP) (Mills, 1991). For higher performance, the precision time protocol (PTP) can be used (IEEE, 2008).

In this analysis, we consider the use case where, during the holdover period, the receiver cannot make use of other timing sources (e.g., Internet connection or other devices) rather than an internal clock and that time obtained from GNSS satellite signal can also not be used as well(despite it may be used as reliable time source under certain circumstances)The aim of this paper is therefore to:

- investigate which are the causes of the misalignment and frequency deviation in clock generators commonly found on the market.
- understand how real-time clock (RTC) specifications are defined and how they relate to the actual operating conditions.
- build a general yet simple model and derive an analytical relation between the clock specification parameters, the time interval between workshop resets, and the OSNMA true accuracy constraints.
- suggest values for clock parameters and frequency of workshop resets that are compatible with available devices for mass market receivers and the OSNMA constraint.

In particular, we focus on two main cases. First, we consider an application where the RTC has the accuracy of a typical GNSS receiver clock. This may be the case of applications that have no strong constraints on energy consumption, such as the smart tachograph, which is installed in a vehicle. We then analyse low-end GNSS receivers that are embedded on portable device with strict limitations on power consumption, and only maintains a very basic, low-energy always on RTC, based on a CMOS oscillator.

The rest of the paper is organized as follows: in Section II, we present the model for clock frequency accuracy and stability, linking these quantities to the requirements of OSNMA; Section III presents the main sources of accuracy loss with focus on those due to temperature variation, the highest contributor to the losses in the commercial clock oscillators. In Section IV we relate the accuracy losses to the OSNMA synchronization requirements and evaluate the performance of clock oscillators found on the market. Lastly, we draw the conclusions on Section V.

## II. SYSTEM MODEL FOR FREQUENCY ACCURACY AND STABILITY

In this Section, we introduce the specification for time and frequency measurements of RTCs. To evaluate the performance of an oscillator two metrics are usually employed (IEEE, 2009):

- the clock frequency accuracy, or normalized difference between the frequency output and its nominal value,
- the clock frequency stability, or normalized instantaneous frequency deviation from its local mean. Usually, frequency stability is distinguished between short-term stability, which is measured in intervals $T_0 < 100$ s, and long-term stability which consider measures with $T_0 > 1$ day.

It is worth to point out that, stability "indicates how well an oscillator can produce the same time or frequency offset over a given time interval" (Lombardi, 2002) hence it does not take account of the actual error between output and nominal value: this means that an oscillator may be stable but not accurate in the long term, e.g. slowly but constantly deviating from the nominal frequency. On the other hand, an oscillator may be unstable but accurate, on average.

To derive a general model, we introduce the following nominal quantities as relative to the output of an ideal clock:

- the (true) reference time, $t$;
- the nominal oscillator frequency output $f_0$;
- the nominal phase (i.e., number of cycles) of the clock at time $t$, $\theta \triangleq f_0 t$.

The output of a RTC can be modelled as (Riley & Howe, 2008)

$$V(t) \triangleq V \sin(2\pi\theta_c(t)) = V \sin(2\pi f_0 T_c(t)) \tag{1}$$

from which we can derive

- the *measured clock phase* at time $t$, $\theta_c(t)$;
- the actual *measured clock time*, $T_c(t) \triangleq \theta_c(t)/f_0$;
- the actual instantaneous oscillator frequency at time $t$, $F_c(t) \triangleq \frac{d}{dt}\theta_c(t)$.

Using the above defined quantities, we can compare the output of any RTC to the output of the ideal clock oscillator, where the reference ideal clock is in practice given by the Galileo system time (GST), and evaluate its performance by computing

- the instantaneous frequency deviation $\Delta F(t) = F_c(t) - f_0$;
- the clock misalignment at time $t$, $\Delta T(t) \triangleq T_c(t) - t$

Moreover, for an interval $T_0$, we also consider

- the average clock frequency during an interval $T_0$ around $t$,

$$\bar{F}_{T_0}(t) \triangleq \frac{1}{T_0} \int_{t-\frac{T_0}{2}}^{t+\frac{T_0}{2}} F_c(\tau)\, d\tau \tag{2}$$

- the instantaneous frequency deviation from its local average over a $T_0$ interval, $\delta F_{T_0}(t) = F_c(t) - \bar{F}_{T_0}(t)$.

Finally, we formally define:

- the clock frequency accuracy as

$$y(t) \triangleq \frac{\Delta F(t)}{f_0}; \tag{3}$$

- the clock frequency stability over $T_0$ as

$$y_s(t) \triangleq \frac{\delta F_{T_0}(t)}{\bar{F}_{T_0}(t)}. \tag{4}$$

Both quantities are typically expressed in *parts per million* (ppm) or *parts per billion* (ppb).

Observe that the following relationship hold between stability and accuracy

$$y(t) = y_s(t) + \frac{\bar{F}_{T_0}(t) - f_0}{f_0} \tag{5}$$

So that, if $|\bar{F}_{t_0}(t) - f_0| \ll f_0 y_s(t)$ stability and accuracy are approximately equal, i.e., $y(t) = y_s(t)$. The possible sources of instability and accuracy loss for our scenarios will be analysed in Section III.

*A. Clock performance requirements for OSNMA*

In this Section we relate the requirements imposed by authentication protocols to clock performance. Still, as pointed out in previous sections, while we will focus on OSNMA, we remark that all the GNSS authentication protocols pose synchronization requirements, thus, our work can be easily translated to target other security mechanism.

The loose time synchronization requirement $T_L$, imposed by OSNMA in (Fernandez-Hernandez, et al., 2020a), states that authenticity of the navigation message received at time $t$ is guaranteed under the assumption

$$|\Delta T(t)| \leq T_L. \tag{6}$$

However, several effects make the clock misalignment grow over time: to bound the clock misalignment, it is envisioned that each device will have a periodic workshop reset. Thus, we assume that

- after a workshop reset at time $t_0$, the next reset is performed at $t_0 + T_R$;
- the calibration performed during each workshop reset brings the misalignment to a negligible value, $|\Delta T(t_0^+)| \ll T_L$, where $t_0^+$ is the instant immediately after the reset; for simplicity, from now on, we will consider $\Delta T(t_0^+) = \Delta T(t_0^+ + T_R) = 0$.

Notice that, with the latter point, we are actually only considering the *phase calibration* error, and not the *frequency calibration* error: we will discuss the contribution of this error on the overall RTC performances on Section III-G. The previous bound (6) can then be restricted to

$$|\Delta T(t)| \leq T_L, \ t \in (t_0, t_0 + T_R) \tag{7}$$

Finally, by using the above definitions, we can relate accuracy and misalignment using the equation

$$\Delta T(t) = \frac{1}{f_0}\left(\theta_c(t) - \theta(t)\right) = \frac{1}{f_0}\int_{t_0}^{t} \Delta F(t)\, d\tau = \int_{t_0}^{t} y(t)d\tau \qquad (8)$$

and hence

$$|\Delta T(t)| \leq \int_{t_0}^{t} |y(t)|d\tau. \qquad (9)$$

which allows to upper bound the clock misalignment at time $t$ in terms of a bound on the clock frequency accuracy along the whole interval from $t_0$ to $t$.

### III. SOURCE OF ACCURACY LOSS FOR RECEIVER CLOCK OSCILLATORS

While there are several technologies available for the realization of clock generators, taking into account performance requirements, size, power consumption, and market price we can restrict our analysis to Quartz crystal oscillators, which are already in use for most GNSS commercial receivers. Indeed, the Quartz crystal oscillators are cheaper than, for instance, atomic clocks and are designed to work at fairly wide operating temperature ranges, spanning from moderate, e.g., $[0, +70°C]$ to wider ranges, e.g., $[-40, +105°C]$.

Still, in energy-saving contexts, CMOS-based oscillators are often used as secondary RTCs: a more precise and energy consuming primary RTC, for instance an OCXO, is used when the device is powered on, while a less precise but less energy consuming secondary RTC to be used when the device is powered off. Then at the turn on, the time is transferred from the less precise clock to the more precise clock. Concerning CMOS oscillators, the most relevant source of inaccuracy are temperature and aging. In Section IV.B, we will analyse the accuracy achieved when using XOs oscillators as primary and CMOS as secondary RTCs.

The main component of any Quartz oscillator is, of course, the Quartz crystal. These crystals are piezoelectric materials and therefore any unwanted stress applied on the crystal will generate an additional voltage, affecting the clock stability. Consequently, the main sources of inaccuracy for a crystal oscillator are

- temperature changes,
- long term aging,
- gravity acceleration,
- other accelerations and vibrations,
- power supply oscillations,
- initial frequency offset after calibration.

More in detail, we distinguish among three classes of crystal oscillators.

**Generic crystal oscillators (XOs)**: XOs are the cheapest oscillators found on the market, but they are also the most susceptible to temperature changes with respect to the calibration conditions (typically, at $25°C$) altering the stability by a factor up to 50ppm.

**Temperature-controlled crystal oscillators (TCXOs):** TCXOs are voltage-controlled oscillators equipped with temperature sensitive electronics that can predict the temperature induced losses and introduce a voltage correction that compensates them; typically they achieve a stability of few ppm, but some more expensive high-precision oscillators can reach $0.1$ppm.

**Oven-controlled crystal oscillators (OCXOs):** OCXOs are crystal oscillators provided with an internal oven that keeps the temperature of the crystal more stable: this allows the oscillator to exhibit performances close to the calibration temperature even in harsher working conditions; this however comes with a cost in terms of size, price, and power consumption.

In Table I we summarize the main characteristics of the three categories, including the typical order of magnitude of accuracy in working conditions. More precise models will be introduced in the next section and specific example values will be given in Section IV.

Taking into account all these factors we choose to restrict our analysis to TCXOs as a good trade-off between stability and cost.

In this Section, we will discuss and model the accuracy losses due to these sources.

*Table 1 Summary of the main quartz crystal oscillator characteristics.*

|      | Power Consumption | Price      | Frequency Accuracy |
|------|-------------------|------------|--------------------|
| XO   | < 1 mW            | 1€ – 10€   | > 10 ppm           |
| TCXO | ≈ 1 mW            | 1€ – 10€   | ≈ 1 ppm            |
| OCXO | > 1 W             | ≫ 10€      | ≈ 0.1 ppm          |

*A. Temperature Changes*

In general, temperature variations change the size of mechanical devices: then, a change in the shape of the crystal may generate additional frequency modes inducing frequency shifts. As listed in datasheets, temperature is often the biggest source of loss in accuracy. First, we will discuss how such loss is reported on the clock specifications. Next, we will briefly describe some of the tests used by the manufactures to establish the bound on accuracy loss.

Datasheets for commercial oscillators typically report the operating temperature range $[x_{\min}, x_{\max}]$ which means that the manufacturers assure that the oscillator will operate properly only if its temperature $x$ lies between $x_{\min}$ and $x_{\max}$.

For what concerns XOs, there is a well-known analytic relationship between temperature and upper bound to frequency accuracy (Valeanu, 2016)

$$Y_{\text{temp}}(x) = A(x - x_0)^2, \tag{10}$$

where $A$ is a parabolic coefficient, and $x_0$ is the calibration temperature, which are typically related to the cut of the quartz crystal itself. Typical values are $A = -0.44$ ppm/°C$^2$ for a calibration temperature of $x_0 = 25$ °C.

Still, to the best of the authors' knowledge, there is no general model relating temperature and accuracy for TCXO or OCXO, i.e., taking into account also the frequency compensations induced by these devices.

Hence, the inaccuracy induced by the temperature changes, $Y_{\text{temp}}(x)$, is bounded

- often by a constant value, $Y_{\text{temp}}$, which represents the maximum frequency accuracy deviation experienced by an oscillator working within the given operating temperature range $[x_{\min}, x_{\max}]$, i.e.,

$$Y_{\text{temp}} = \max_{x \in [x_{\min}, x_{\max}]} |Y_{\text{temp}}(x)|, \tag{11}$$

- less frequently, by a linear function of the difference between $x_0$ and the working temperature condition.

$$Y_{\text{temp}}(x) = A'|x - x_0|. \tag{12}$$

The slope $A'$ is measured in ppb/°C.

In the former case TCXOs operating in wider temperature ranges usually exhibit a higher (maximum) frequency deviation. Still, it is worth to point out that in general the stability of an oscillator may not improve if used in a narrower temperature range; it is more appropriate to identify a target operating temperature range and only compare oscillators working in that range.

While the uncertainty may vary significantly, especially from cheaper to more expensive oscillators, we can consider as typical values $Y_{\text{temp}} = 0.5$ ppm for the range $[0, +70°C]$ and $Y_{\text{temp}} = 2$ ppm for wider range oscillators $[-40, +105\ °C]$ (e.g., see We report in Table 2).

*C. Long Term Aging*

Long term aging has a significant impact on the clock frequency accuracy and may affect the device even when it is not used for a long time. It represents the losses originated by multiple causes, e.g., relaxation of mechanical stress induced by the structure to the Quartz crystal, and migration of impurities within the crystal itself (JEDEC STANDARD, 2000). A critical aspect of this loss in accuracy induced by aging is that it is time-variant: in fact, the accuracy loss introduced by aging accumulates in time, with subsequently decreasing increments, yielding a total uncertainty that increases sub linearly with time. In some cases, after several years the frequency deviation due to long-term aging may even change sign (Riley & Howe, 2008).

We model the accuracy loss due to aging as $y_{\text{age}}(t)$. Since the aging loss increment decreases over time, the loss measured before the first reset is typically significantly higher than the loss measured at subsequent resets.

In general, for a reset interval $T_R$, we are interested in learning the corresponding long-term aging accuracy bound at the first reset $Y_{\text{age}}(T_R)$. However, in most datasheets we will not find such value: typically, they report a single value $Y_{\text{age}}(T_{\text{data}})$ for $T_{\text{data}} = 1$ year or $T_{\text{data}} = 10$ years. Hence, we need to derive a general bound from such values.

The effect of long term aging for both TCXOs and OCXOs was investigated in (Filler & Vig, 1993) measuring the accuracies of different oscillators for several years: in particular, the frequency offset of a TCXO that went through temperature cycles between $x^+ = +60°C$ and $x^- = -40°C$ was measured, by taking as reference a caesium oscillator at the same nominal frequency $f_0$. From the collected frequency deviation measurements $\Delta \hat{F}(t)$ and

$$\hat{Y}_{\text{age}}(t) = \frac{\Delta \hat{F}(t)}{f_0}, \tag{13}$$

both a linear and a logarithmic model where fitted, with parameters described respectively by

$$Y_{\text{lin}}(t) = At + C, \tag{14}$$

and

$$Y_{\text{log}}(t) = B \ln(Dt + 1). \tag{15}$$

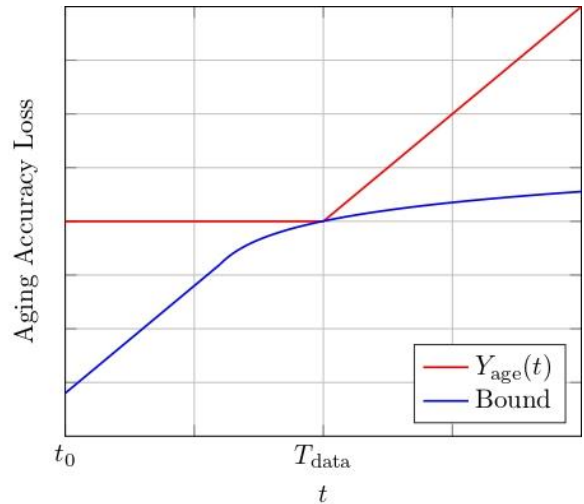

Fig. 2. Graphical representation of the model (16): upperbound (red) versus estimated model of (Filler & Vig, 1993) (blue).

The study concluded that:
- The logarithmic fit is better suited for long-term measurements.
- The linear fit is better suited for initial measurements (i.e., less than 30 days), while it becomes a loose upper-bound for longer times.

Since we are interested in establishing a prudential upper bound rather than a precise estimate, we use a constant upper bound for $t < t_0 + T_{\text{data}}$ and a linear upper bound for $t > t_0 + T_{\text{data}}$

$$\begin{cases} |Y_{\text{age}}(t)| \leq Y_{\text{age}}(T_{\text{data}}), & \forall t \in (t_0, t_0 + T_{\text{data}}), \\ |Y_{\text{age}}(t)| \leq Y_{\text{age}}(T_{\text{data}}) \frac{t - t_0}{T_{\text{data}}}, & \forall t \in (t_0 + T_{\text{data}}, t_0 + T_R). \end{cases} \tag{16}$$

Clearly, the linear bound overestimates the aging loss: it would be ideal then to find manufacturer data that provides an estimation of the effect of aging as close as possible to the reset period $T_R$ of the device. A customary value of the long-term aging for a TCXO is $Y_{\text{age}}(T_{\text{data}}) = 1$ ppm at $T_{\text{data}} = 1$ year, so, for instance, if we choose $T_R = 2$ years we can upper bound $|y_{\text{age}}(t_0 + T_R)| \leq Y_{\text{age}}(t_0 + T_R) = 2$ ppm.

Fig. 2 reports the comparison between the aging loss fits, obtained from (14) and (15), and the upper bound (16), as a function of time $t$.

*D. Gravity*

Gravity applies a constant stress on the crystal, unlike shocks and vibrations that can be reduced using a proper damping packaging (Hewlet Packard, 1997). It is then easy to notice that this effect is relevant only in those scenarios where gravity is very different from the one at calibration, hence it can be neglected in an automotive scenario on ground.

*E. Acceleration and Vibrations*

The effects on the clock stability due to acceleration and vibrations are not always reported on datasheets; when available, this effect is quantitatively expressed as an upper bound proportional to the acceleration $a$

$$Y_{\text{vib}}(a) = Ka, \tag{17}$$

with $K$ expressed in ppb/$g$ with $g$ denoting gravitational acceleration. This uncertainty is because the vibrations applied to the device, such as the ones typical of an automotive scenario, cause stress on the crystal.

While it is hard to correctly measure this uncertainty, we find that a typical value is $K = 0.1$ ppb/$g$ (Vig, 2016), hence we can reasonably neglect this effect, as well, since typical accelerations are in the range of fractions of $g$.

*F. Power Supply Variations*

As for any electronic component, oscillations on the supply voltage $V_{\text{CC}}(t)$ of the device may be a source of instability. Denoting the nominal voltage $V_{\text{CC}}$ and the oscillations by $\Delta V_{\text{CC}}(t) = |V_{\text{CC}}(t) - V_{\text{CC}}|$ a typical bound for the stability associated with this source is bounded by $Y_{\text{supp}} = 0.1$ppm for $|\Delta V_{\text{CC}}(t)/V_{CC}| \leq 5\%$ (Vig, 2016). This problem can be solved by pairing the oscillator with a voltage regulator. We will assume that the supply voltage is kept constant and neglect this source.

*G. Initial Frequency Offset after Calibration*

The calibration and reset processes are in general prone to errors: while the phase calibration error is neglectable in this framework, since it is a constant offset and it is a much smaller contribution which respect to the other losses, the accuracy loss due to frequency calibration error, $f_{\text{calib}}$, grows over time. An off-the-shelf oscillator has a frequency calibration error $f_{\text{calib}}$ limited by frequency tolerance $f_{\text{tol}}$, with a typical value of $1 - 2$ ppm. To mitigate this error, it is possible to perform a precise calibration, trying to synchronize the RTC with the actual $f_0$, e.g., by using an atomic clock. Hence, we bound the accuracy loss due to the initial frequency calibration error as

$$y_{\text{calib}} = \frac{f_{\text{calib}}}{f_0} \leq Y_{\text{calib}}, \tag{18}$$

where $Y_{\text{calib}}$ is an a priori value decided either by system design or during the calibration process itself. Still, we assume that the calibration process to be perfect or, equivalently, we assume the errors due to calibration to be much lower than the errors due to the major source of errors, i.e., temperature and aging.

*H. Turn-Off/Turn-On Bias*

An additional effect to be considered is the bias induced by repeated turn on and turn off operations. This happens when, for energy saving purposes, the clock is turned off, and then turned on when needed. This typically induces non predictable errors (IEEE, 2008).

Thus, in our analysis we will consider two scenarios:

- a high-power scenario, where we assume that no limitation is imposed on the energy consumption of the clock, thus there is no need for turning it on/off, and
- a power-limited scenario, where the clock may be turned off to save energy, relying on a less precise clock for the turn off period. Next, at start up, time is transferred back to the more precise clock.

The first scenario may model several cases such as when using the clock in static conditions, e.g., on a base station for scheduling purposes in a 5G network. It may also be considered for the smart tachograph, installed in a truck which can afford to keep the same clock on. The latter targets instead scenarios where the clock is embedded on a device that has a constraint on the energy consumption, e.g., an autonomous vehicle or an IoT device.

Indeed, in the first scenario we can naturally neglect the turn-off/turn-on bias as the clock can be always on. in the second scenario, we will tackle the worst-case analysis, where the performance of the system is determined solely by the performance of the secondary, less performing clock, i.e., as if the primary clock was off most of the time.

*I. Overall accuracy*

In general, the cross-correlation between the uncertainties is unknown; we can only consider the worst-case scenario where the total uncertainty is bounded by the sum of the bounds on the single uncertainties. This approach is confirmed by (MIL-PRF-55310F, 2018). We underline that this choice represents a prudential and conservative approach that it is likely to yield a loose bound.

Hence, after the workshop reset at time $t_0$,

$$y(t) = y_{\text{temp}}(t) + y_{\text{age}}(t) + y_{\text{grav}}(t) + y_{\text{vib}}(t) + y_{\text{supp}}(t) + y_{\text{calib}}, \qquad t \in (t_0,\ t_0 + T_R), \tag{19}$$

and

$$|y(t)| < Y = Y_{\text{temp}} + Y_{\text{age}}(t),\ t \in (t_0,\ t_0 + T_R), \tag{20}$$

where

- $y_{\text{temp}}(t)$ is bounded by $|y_{\text{temp}}(t)| \leq Y_{\text{temp}}, \forall t \in (t_0, t_0 + T_R)$;
- $y_{\text{calib}}$ the accuracy loss due to the residual frequency offset after the clock reset, bounded by $|y_{\text{calib}}| < Y_{\text{calib}}$, which will be neglected by choosing a strict enough $Y_{\text{calib}}$;
- $y_{\text{age}}(t)$ is due to aging effect and is bounded as in (16);
- $y_{\text{grav}}(t)$, $y_{\text{vib}}(t)$, and $y_{\text{supp}}(t)$ can be neglected, as discussed on the dedicated sections.

Finally, we will consider two separate scenarios, as described in III.H. Notice for the latter scenario many of the manufacturers already report the accuracy including both temperature and aging. Indeed, for CMOS the most relevant effect is temperature, with $Y_{\text{temp}}$ typically one order of magnitude larger than $Y_{\text{age}}$. Thus, for CMOS, we will use the following approximation

$$|\Delta T(t)| \leq Y_{\text{temp}} T_R + \frac{1}{2} Y_{\text{age}}(T_{\text{data}}) \left(T_{\text{data}} + \frac{T_R^2}{T_{\text{data}}}\right) \approx Y_{\text{temp}} T_R \triangleq B(T_R). \tag{24}$$

## IV. EVALUATION OF THE BOUND ON THE CLOCK ERROR MISALIGNMENT

Recalling the bounds in Section II, we can relate accuracy and time misalignment by (9), and combining it with (20),

$$|\Delta T(t)| \leq \int_{t_0}^{t} \left(Y_{\text{temp}} + Y_{\text{age}}(t)\right) dt = (t - t_0) Y_{\text{temp}} + \frac{1}{2}\left(T_{\text{data}} + \frac{(t-t_0)^2}{T_{\text{data}}}\right) Y_{\text{age}}(T_{\text{data}}). \tag{21}$$

As (21) increases with $t$, we introduce upper the bound on the clock error $\forall t \in (t_0, t_0 + T_R)$ as

$$|\Delta T(t)| \leq Y_{\text{temp}} T_R + \frac{1}{2} Y_{\text{age}}(T_{\text{data}}) \left(T_{\text{data}} + \frac{T_R^2}{T_{\text{data}}}\right) \triangleq B(T_R), \tag{22}$$

where we used the assumption that at time $t_0$, the RTC is perfectly synchronized (and syntonised) with the reference time. The bound $B(T_R)$ can be also interpreted as the maximum misalignment experienced by the oscillator in the worst-case scenario between the first calibration and the next reset, i.e., $\forall t \in (t_0, t_0 + T_R)$.

As per the OSNMA Receiver Guidelines (European Union, 2024), we consider loose time synchronization requirements of $T_L = 165$ s or $T_L = 15$ s and $T_R = 2$ years. The value $T_L = 165$ is based on the 'SLOW MAC', from the OSNMA specification (European Union, 2023). In the next, we consider separately the high power and the limited power scenarios.

Considering the results of Section III, we can state that the oscillator under test is suitable for OSNMA if

$$B(T_R) \leq T_L,\ \forall t \in (t_0, t_0 + T_R). \tag{23}$$

We can then exploit this expression to compare the requirements for different commercial oscillator by manufacturers and model. Notice that we assumed that all the devices will have to undergo a calibration process such that $Y_{\text{calib}} \ll Y_{\text{temp}}$ thus we can neglect this term in the actual calculations.

We report in Table 2

- the values of the misalignment bound, $B(T_R)$ for a workshop reset period $T_R = 2$ years;
- the maximum reset period $T_{R,\max}$ such that $B(T_{R,\max}) \leq T_L$, with a loose time synchronization requirement $T_L = (165\ s, 15s)$.

as computed form the specs found in the datasheets. We consider TCXO designed for GNSS receivers, with operating frequency $f_0$ not higher than 52 MHz. Since we are considering oscillators for a device that will operate in the automotive scenario, an environment where the operating temperature may change a lot, we considered as target operating temperature range $[x_{\min}, x_{\max}] = [-20°C, +85 °C]$.

Moreover, notice that once an oscillator is chosen and fixing the loose synchronization requirement $T_L$ and a revision period $T_R$ given by the service provider, if inequality (21) is satisfied with ample margin, it is possible to relax the constraint on the calibration accuracy $Y_{\text{calib}}$ during the factory reset, potentially speeding up the process.

### A: High Power Scenario

As outlined in Section III.H, we consider now the scenario where the device is constantly powered on (or holds sufficient energy) so that the primary precise clock is never turned off for between $t_0$ and $t_0 + T_R$.

We report in Table 2 both the bounds values $B(T_R)$ and $T_{R,\max}$ computed using several RTCs datasheet specs.

Table 2: Bound values $B(T_R)$ and reset period $T_{R,\max}$, for loose time synchronization requirements $T_L = 15$ s and $T_L = 165$ s, derived from the TCXO's datasheets.

| Manufacturer | Model | Oper. Temp [°C] $x_{\min}$ | Oper. Temp [°C] $x_{\max}$ | $Y_{\text{temp}}$ | $Y_{\text{age}}$(1 year) | $B(T_R)$ [s] | $\dfrac{T_{R,\max}}{T_L = 15\,s}$ [days] | $\dfrac{T_{R,\max}}{T_L = 165\,s}$ [years] |
|---|---|---|---|---|---|---|---|---|
| SEIKO EPSON | TG-5035CJ | −40 | +105 | 0.5 ppm | 1 ppm | 110.38 | 115.74 | 2.62 |
| | TG2016SMN | −40 | +90 | 0.5 ppm | 0.5 ppm | 70.96 | 173.61 | 3.57 |
| | TG2016SLN | −40 | +85 | 0.5 ppm | 1 ppm | 110.38 | 115.74 | 2.62 |
| | TG-5006CJ | −30 | +85 | 0.5 ppm | 1 ppm | 110.38 | 115.74 | 2.62 |
| | TG2016SKA | −40 | +105 | 0.5 ppm | 1 ppm | 110.38 | 115.74 | 2.62 |
| VECTRON | VT-803 | −40 | +85 | 1 ppm | 0.5 ppm | 102.49 | 115.74 | 2.89 |
| | VT-706 | −40 | +85 | 0.5 ppm | 1 ppm | 110.38 | 115.74 | 2.62 |
| | VT-702 | −40 | +85 | 0.5 ppm | 1 ppm | 110.38 | 115.74 | 2.62 |
| | VT-804 | −40 | +85 | 2 ppm | 1 ppm | 204.98 | 57.87 | 1.67 |
| NDK | NT2520SE | −40 | +105 | 0.5 ppm | 1 ppm | 110.38 | 115.74 | 2.62 |
| | NT1612AA | −30 | +85 | 0.5 ppm | 1 ppm | 110.38 | 115.74 | 2.62 |
| | NT1612AJA | −30 | +85 | 0.5 ppm | 1 ppm | 110.38 | 115.74 | 2.62 |
| | NT2016SA | −30 | +85 | 0.5 ppm | 1 ppm | 110.38 | 115.74 | 2.62 |
| Maxim Integrated | DS3231 | −40 | +85 | 3.5 ppm | 1 ppm | 299.59 | 38.58 | 1.16 |
| Micro Crystal Switzerland | RV-8803-C7 | −40 | +85 | 3 ppm | 3 ppm | 425.73 | 28.94 | 0.87 |

### B. Power Limited Scenario

Now, we consider the power limited scenario discussed in Section III.H, modelled following a worst-case analysis, i.e., assuming that the accuracy depends solely on the secondary clock. In particular, we considered RTC CMOS clocks. Table 3 reports bound values $B(T_R)$ and reset periods $T_{R,\max}$, for loose time synchronization requirements $T_L = 15$ s and $T_L = 165$ s.

Table 3 reports the bounds values $B(T_R)$ and $T_{R,max}$ computed using several RTCs datasheet specifications, for $T_L = 15$ s and $T_L = 165$ s. Indeed, when compared with the no power limit scenario (Table 2), we see that a device using only CMOS RTC frequent factory reset are needed. For instance, the reset period $T_R$ spans over few days for $T_L = 15$ s and $T_R$ and over 1 to 4 months for $T_L = 165$ s, when using CMOS RTC. On the other hand, in the high power scenario, where the TCXO is used for the whole period, we get $T_R > 1$ month, even for the stricter requirement $T_L = 15$ s.

Table 3 Bound values $B(T_R)$ and reset period $T_{R,max}$, for loose time synchronization requirements $T_L = 15$ s and $T_L = 165$ s, derived from the CMOS RTC datasheets.

| Manufacturer | Model | Oper. Temp [°C] $x_{min}$ | Oper. Temp [°C] $x_{max}$ | $Y_{temp}$ | $Y_{age}(1\text{ year})$ | $B(T_R)$ [min] | $T_{R,max}$ $T_L = 15\text{ s}$ [days] | $T_{R,max}$ $T_L = 165\text{ s}$ [days] |
|---|---|---|---|---|---|---|---|---|
| MICREL | DSC1003 | −40 | +105 | 10 ppm | 5 ppm | 17.08 | 11.57 | 127.31 |
| | | | | 25 ppm | | 32.85 | 5.78 | 63.66 |
| | | | | 50 ppm | | 59.13 | 3.16 | 34.72 |
| TXC[2] | 7X | −40 | +85 | 20 ppm | 3 ppm | 21.02 | 8.68 | 95.49 |
| | | | | 25 ppm | | 26.28 | 6.94 | 76.39 |
| | | | | 50 ppm | | 52.56 | 3.47 | 38.19 |
| | 7C | −40 | +85 | 15 ppm | 3 ppm | 15.76 | 11.57 | 127.31 |
| | | | | 25 ppm | | 26.28 | 6.94 | 76.39 |
| | | | | 30 ppm | | 31.52 | 5.79 | 63.66 |
| | | | | 50 ppm | | 52.56 | 3.47 | 38.19 |
| ECS[1] | ECS-327ATQ2016MV | −40 | +125 | 50ppm | / | 52.56 | 3.47 | 38.19 |

## V. CONCLUSIONS

GNSS authentication services such as OSNMA require loose time synchronization between transmitter and receiver to ensure the authenticity of the signals. Due to the frequency drift of RTCs, it is necessary to pre-establish a reset period ($T_R$) to bound the misalignment between transmitter and receiver clocks, which needs to be lower than the synchronization requirement $T_L$.

In this paper, we have first investigated which are the causes of the clock misalignment and clock frequency deviation for clock generators, discussing their relevance in the automotive scenario and defined a general relationship between workshop reset period, $T_R$, the requirement $T_L$, and the clock specifications commonly found in the datasheets. Next, we examined several mass-market temperature-controlled crystal oscillators (TCXOs) and CMOS-based oscillators datasheets evaluating their performance in terms of worst-case scenario accuracy bound $B(T_R)$.

We underline that this choice represents a prudential approach that yields a conservative bound with a very high probability, and looks like a reasonable solution given the lack of a constant statistical model. We concluded that most of the devices were able to satisfy the constraints $B(T_R) \leq T_L = 165$ s with a workshop reset period $T_R$ = 2 years when considering a non-power limited scenario, where the TCXO-based RTC is turned on all the time.

Further work may include an experimental activity involving the test of real clocks, to derive a more refined model for accuracy and stability, eventually including also the statistical distributions for the different error contributions.

---

[2] The accuracy due to temperature includes also the effects of supply voltage, load and aging.

**Declarations**

*Ethics approval and consent to participate*

Not Applicable.

*Availability of data and materials*

Not Applicable.

*Funding*

Not Applicable.

**References**

Allan, D. W. (1966). Statistics of atomic frequency standards. *Proceedings of the IEEE*, 221-230. doi:10.1109/PROC.1966.4634

Anderson, J. M., Carroll, K. L., DeVilbiss, N. P., Gillis, J. T., Hinks, J. C., O'Hanlon, B. W., . . . Yazdi, R. A. (2017). Chips-Message Robust Authentication (Chimera) for GPS Civilian Signals. *International Technical Meeting of the Satellite Division of The Institute of Navigation (ION GNSS+ 2017)* (pp. 2388-2416). Portland, Oregon: ION. doi:10.33012/2017.15206

Baldini, G., Sportiello, L., Chiaramello, M., & Mahieu, V. (2018,). Regulated applications for the road transportation infrastructure: The case study of the smart tachograph in the European Union. *International Journal of Critical Infrastructure Protection2018,, 21*, 3-21. doi:10.1016/j.ijcip.2018.02.001

Cartright, J. (2008). *Aging Performance in Crystals.* Retrieved from http://www.conwin.com/pdfs/aging_perf_crystals.pdf

Cucchi, L., Damy, S., Paonni, M., Nicola, M., Gamba, M. T., Motella, B., & Fernandez-Hernandez, I. (2021). Assessing Galileo OSNMA Under Different User Environments by Means of a Multi-Purpose Test Bench, Including a Software-defined GNSS Receiver. *International Technical Meeting of the Satellite Division of The Institute of Navigation (ION GNSS+ 2021)* (pp. 3653-3667). St. Louis, Missouri: ION. doi:https://doi.org/10.33012/2021.17966

Djenouri, D., & Bagaa, M. (2016). Synchronization Protocols and Implementation Issues in Wireless Sensor Networks: A Review. *IEEE Systems Journal* , 617 - 627. doi:10.1109/JSYST.2014.2360460

European Union. (2023). *Galileo open service navigation message authentication (OSNMA) signal-in-space interface control document (SIS ICD).* Retrieved March 2024, from https://www.gsc-europa.eu/sites/default/files/sites/all/files/Galileo_OSNMA_SIS_ICD_v1.1.pdf

European Union. (2024). *Galileo open service navigation message authentication (OSNMA) receiver guidelines.* Retrieved March 2024, from https://www.gsc-europa.eu/sites/default/files/sites/all/files/Galileo_OSNMA_Receiver_Guidelines_v1.3.pdf

EUSPA. (2024). *PATROL: Position Authenticated Tachograph foR OSNMA Launch*. Retrieved March 2024, from https://www.patrol-osnma.eu/

Fernández-Hernández, I., Rijmen, V., Seco-Granados, G., Simon, J., Rodríguez, I., & Calle, J. D. (2016, April). A Navigation Message Authentication Proposal for the Galileo Open Service. *NAVIGATION, Journal of the Institute of Navigation, 63*, 85-102. doi:10.1002/navi.125

Fernandez-Hernandez, I., Walter, T., Neish, A., & O'Driscoll, C. (2020). Independent Time Synchronization for Resilient GNSS Receivers. *International Technical Meeting of The Institute of Navigation* (pp. 964 - 978). San Diego, California: ION. doi:10.33012/2020.17190

Fernandez-Hernandez, I., Winkel, J., O'Driscoll, C., Cancela, S., Terris-Gallego, R., Lopez-Salcedo, J. A., . . . de Blas, J. (2023). Semiassisted Signal Authentication for Galileo: Proof of Concept and Results. *IEEE Transactions on Aerospace and Electronic Systems*, 4393 - 4404. doi:10.1109/TAES.2023.3243587

Filler, R., & Vig, J. (1993). Long-term aging of oscillators. *IEEE Transactions on Ultrasonics, Ferroelectrics, and Frequency Control, 40*(4), 387 - 394. doi: 10.1109/58.251287

Galleani, L., Sacerdote, L., Tavella, P., & Zucca, C. (2003). A mathematical model for the atomic. *Metrologia, 40*(3). doi:10.1088/0026-1394/40/3/305

Giorgi, G. (2015). An Event-Based Kalman Filter for Clock Synchronization. *IEEE Transactions on Instrumentation and Measurement*, 449 - 457. doi:10.1109/TIM.2014.2340631

Hewlet Packard. (1997). *Fundamentals of Quartz Oscillators: Application Note 200-2.* Retrieved from http://leapsecond.com/hpan/an200-2.pdf

IEEE. (2008). IEEE Standard for a Precision Clock Synchronization Protocol for Networked Measurement and Control Systems. doi:https://doi.org/10.1109/IEEESTD.2008.4579760

IEEE. (2009). IEEE Standard Definitions of Physical Quantities for Fundamental Frequency and Time Metrology---Random Instabilities. doi:10.1109/IEEESTD.2008.4797525

JEDEC STANDARD. (2000). JESD22-A104-B. *Temperature Cycling*.

Lombardi, M. A. (2002). Fundamentals of Time and Frequency. In R. H. Bishop, *The Mechatronics Handbook.* CRC Press.

Mills, D. L. (1991). Internet Time Synchronization: the Network Time Protocol. *IEEE Transactions on Communications*, 1482-1493. doi:10.1109/26.103043

MIL-PRF-55310F. (2018). *Performance Specification: Oscillator Crystal Controlled, General Specificification for*.

MIL-STD-883. (2004). *Test Method Standard, Microcircuits*.

Perrig, A., Canetti, R., Tygar, J., & Song, D. (2002, November ). The TESLA broadcast authentication protocol. *RSA CryptoBytes, 5*, 2-13. doi:10.1007/978-1-4615-0229-6_3

Prakash, R., & Kendall, N. (2010, Apr.). Time synchronization in wireless sensor networks: A survey. *International Journal of UbiComp, 2*(1), 92-102. doi:10.5121/iju.2010.1206

Riley, W., & Howe, D. A. (2008). *Handbook of Frequency Stability Analysis.* NIST.

Tirado-Andrés, F., & Araujo, A. (2019). Performance of clock sources and their influence on time synchronization in wireless sensor networks. *International Journal of Distributed Sensor Networks, 15*(9). doi:10.1177/1550147719879372

Valeanu, A. (2016). *Temperature Compensation of a Tuning Fork Crystal Based on MCP7941X.* Retrieved from https://ww1.microchip.com/downloads/en/Appnotes/00001413B.pdf

Vig, J. R. (2016). Quartz Crystal Resonators and Oscillators for Frequency Control and Timing Applications: A Tutorial. *IEEE International Frequency Control Symposium Tutorials*.